# Catalytic sub-surface etching of nanoscale channels in graphite

Maya Lukas[1,2,*], Velimir Meded[1,3], Aravind Vijayaraghavan[1,4], Li Song[5,6], Pulickel M. Ajayan[6], Karin Fink[1], Wolfgang Wenzel[1], Ralph Krupke[1,2,7]

[1] Karlsruhe Institute of Technology (KIT), Institute of Nanotechnology, D-76021 Karlsruhe, Germany

[2] DFG Center for Functional Nanostructures (CFN), D-76031 Karlsruhe, Germany

[3] Karlsruhe Institute of Technology (KIT), Steinbuch Centre for Computing, D-76021 Karlsruhe, Germany

[4] School of Computer Science, The University of Manchester, Manchester M13 9PL, United Kingdom

[5] Research Center for Exotic Nanocarbons, Shinshu University, Nagano 380-8553, Japan

[6] Department of Mechanical Engineering & Materials Science, Rice University, Houston, TX 77005, USA

[7] Department of Materials and Earth Sciences, Technische Universität Darmstadt, D-64287 Darmstadt, Germany

[*] email: maya.lukas@kit.edu

**Catalytic hydrogenation of graphite has recently attracted renewed attention, as a route for nano-patterning of graphene and to produce graphene nano-ribbons. These reports show that metallic nanoparticles etch surface layers of graphite[1-8], or graphene[9-10] anisotropically along the crystallographic zigzag <11-20> or armchair <1010> directions. On graphene the etching direction can be influenced by external magnetic fields or the substrate[11-13]. Here we report the sub-surface etching of highly oriented pyrolytic graphite (HOPG) by Ni nanoparticles, to form a network of tunnels, as seen by SEM and STM. In this new nanoporous form of graphite, the top layers bend inward on top of the tunnels, while their local density of states remains fundamentally unchanged. Engineered nanoporous tunnel networks in graphite allow further chemical modification and may find applications in storage or sensing.**

Since the 1970s it is known that metallic nanoparticles can catalyze local hydrogenation of graphite, which results in straight nanoscale channels along the crystallographic high symmetry directions intersecting by integer multiples of 30°[1-10]. To date, clear evidence has been reported only for open channels on the surface of graphite, whereas sub-surface etching remains uncertain [4,5,8]. In this work HOPG was etched by the two-step procedure described previously[7]: After a controlled oxidation step to introduce nucleation sites, Ni nanoparticles were deposited and the sample was subsequently exposed to high temperature in H atmosphere to enable catalytic etching (see methods for details). The SEM image of such a sample shows etch-channels that are mainly oriented along a few preferred directions (Fig. 1a). On closer inspection two different types of channels can be distinguished: The first type (one is marked by a blue arrow in Fig. 1a) appears brighter directly at the edge. The other type (red arrow) does not show this effect. This difference is best seen in the intensity profile (Fig. 1b) along the white line in Fig. 1a. In SEM imaging it is a well-known phenomenon that edges within the scattering cloud of the electron beam give a higher secondary-electron (SE) yield and appear brighter than plane surface parts[14]. Hence channels with bright edges indicate open channels (trenches). The missing of enhanced SE emission in the second type of channels is not dependent on the channel width. It, therefore, cannot be assigned to resolution limits of the apparatus, but implies a different channel morphology.

In STM investigations of the same sample we also observe two different channel types (Fig. 1e). The height profile (Fig. 1f) along the white trace in Fig. 1e illustrates the differences: Some channels, marked by blue arrow, are imaged in the same way as step edges of HOPG: The apparent-height $\Delta z$ shows an enhancement near a step-edge of a monolayer (ML) or multiple ML height. The enhancement is always found on the upper terrace and is due to an altered electronic structure[15-21] or backfolding of the sheet[15,22,23] near the edge. This channel corresponds to a trench where the top layers of graphite have been etched away by a Ni particle.

A second type of channel (red arrow in Fig. 1e) does not show this edge enhancement and was investigated in detail: We find no reconstruction of graphite or height-enhancement near the step edge for the channel in Fig. 2a, as is clearly seen in the height profiles along the traces 1,2 which are plotted in Fig. 2b. Furthermore, the channel-depth with 45-50 pm is by far too small for a ML step on graphite ($\Delta z$=350 pm). Fig. 2c shows the same region with atomic resolution. It is obvious that the periodic structure of the top graphene layer is unperturbed throughout the channel´s area. The height profiles along traces 3,4 (Fig 2d) can be fitted by a cosine function of periodicity 0.247 nm superimposed onto a third-grade polynomial (describing a flexed beam). We therefore conclude that the surface sheet of graphite is intact, but appears to be sagged down along a trace of similar orientation and width as an etch-trench. All sagged traces of this kind have a connection to etch pits, step edges or other etch-trenches. In Fig. 2a this is seen at the top of the image. We thus conclude that the etching process is not limited to the surface, but also takes place in the material below the surface, resulting in tunnels. It might be argued that a very blunt tip would not resolve the depth and the abrupt height change of a narrow ML trench. However, at a ML step-edge the atomic positions change laterally by 71pm and the positions described by the maxima of the sine are shifted by half a period. While the lateral shift might be difficult to recognize, the π-shift in the sine is clearly missing. For a deeper trench with the maxima in phase again atomic resolution becomes impossible with the tip more than a ML further away from the surface. The perfect fit of the sine function beside and on the channel indicates that the top sheet is intact and that the tunnel is covered by at least two layers of graphene (see SI for details). The apparent sagging above the tunnel is not necessarily due to sample topography as will be discussed below.

The existence of etch trenches and tunnels explains the findings in the SEM images (Fig. 1a,c,d) straight away. The contrast enhancement due to increased SE emission from the edge of a trench (blue arrows) cannot occur for a tunnel´s edge buried inside the bulk material (red arrows). Reduced SE generation in the tunnel (no material to scatter) accounts for the effect that tunnels appear darker than the surroundings. The differences can be seen very clearly in Fig. 1c and d where the material above a tunnel is partially removed.

In contrast, the interpretation of the STM-images of tunnels is not straightforward because the tunneling current depends on the tip-sample distance as well as on the local density-of-states (DOS) of the sample. Therefore, the apparent sagging of the graphene layers covering the tunnels as observed in STM may have different causes: (i) graphene layers covering the tunnel may form a sagging ceiling, or (ii) the DOS is decreasing towards the centerline of the tunnel, or (iii) a mixture of both effects.

To gain better understanding of these phenomena we performed semi-empirical quantum chemical calculations using the modified neglect of diatomic differential overlap method[24]. We modeled the tunnel as a stack of several graphene sheets, periodic in the zigzag <11-20> and armchair <10-10> directions, in which a tunnel running along <10-10> was created in the center (see methods). The sidewall of the tunnel corresponds to a straight armchair-edge and was saturated with hydrogen atoms. The tunnel was varied having one, two, or three covering layers. For a single cover-layer we also performed calculations of a wider tunnel.

For these models we performed an energy minimization, permitting all atoms to follow their local gradient until equilibrium was reached. The relaxed structures of lowest energy are shown in Fig. 3 a-d. For all models a downward bending of the top layers was observed. For tunnels of the same width (a-c) the deflection $\Delta z$ of the topmost carbon sheet out of its bulk position decreases from 180 pm for one cover-layer to 51pm for 3 cover-layers. For a single cover-layer, we observed much stronger bending for the wider tunnel, $\Delta z=280$ pm (d).

From these calculations we learn that carbon atoms forming the top of the tunnel will interact with the sheet edges at the side of the tunnel by van-der-Waals (vdW) interaction, pulling the layer inward. If two or three layers are bent, roughly two or three times as many bonds have to be stretched respectively, resulting in less bending as more layers are added.

To investigate a possible charge rearrangement we additionally performed density functional (DFT) single point calculations[25] for a whole graphene layer spanning a layer with gap beneath, in either straight (Fig. 3e) or bent (Fig. 3f) conformations. We find no significant overlap between the DOS of the sheets for either of these conformations. The DOS also shows no shift of electrons towards or away from the atoms above the tunnel as compared to the atoms at the sides. Therefore no essential charge redistribution occurs within the sheet above the tunnel, an effect that is likely to persist if the tunnel is covered by more than one layer of graphene.

The findings indicate that the depression in the STM image is caused by a sagging of the graphene sheets on top of a tunnel covered by few-layer graphene. We also note that interactions between the surface and the tip would pull the top sheet towards the tip[26-28], therefore counteracting the depression we are observing. Since the tip is scanning at rather large distance and we never observe upward bending at our feedback parameters these forces must be smaller than those within the system.

Assuming a spherical shape of the nanoparticle[4], a tunnel height in the order of magnitude of its width can be inferred. In Fig. 2 the width can be determined as approximately 12 $|\hat{u}_{<11-20>}|$ shortest distance across the tunnel, corresponding to a height of 8 cut layers. The dimensions of the observed tunnel are thus of the same order of size as the tunnels investigated in the models above. In our images we find tunnels with comparable width but different bending. We assign this to different number of topping sheets. Indentations of several Å depth are only observed on wider tunnels or larger irregularly shaped etching "caves".

It is interesting to note, that a similar appearance in STM as found for our tunnels and caves was recently reported for graphene grown over Cu vacancies[28]. Though the interaction with the underlying Cu is necessarily different, resulting in a different depth of the indentation, the general appearance with continuous atomic corrugation, smoothly varying height and no edge enhancement is similar.

As a last proof for the existence of tunnels and to estimate if there is a limit, how deep below the surface tunnels can be etched, we performed a triangulation experiment in SEM (details in SI). Fig. 4a shows an SEM image with a trench (marked blue) and a tunnel (marked red). After tilting the sample by 45° (Fig. 4b), the distance of small surface particles (labeled 1-5) relative to the tunnel edge has changed due to projection. From the distance change a depth of 75nm below the surface can be calculated for this tunnel. This is on the order of the theoretical SE escape-depth in graphite (100 nm)[29]. The depth at which tunnels can be etched appears to be limited by no other restrictions than the depth of etch pits or the stepped structure of the surface, as tunnels are found at all depth levels, from few layers below the surface (STM) down to the probing-limit of SEM.

As already reported for etching trenches, we see no indication for tunnels with a direction component perpendicular to the layers, i.e. cutting through layers, in our experiments. In general trenches and tunnels start at step edges and in etch pits, the particles act on the terminations at the side of the sheets, thus etching parallel to the layers. From STM images it is found, that the number of layers on top of a tunnel can vary along its length, by one or two layers. Also the etching direction is- on the atomic scale- not strictly straight. We attribute this to a reshaping of the particle on the atomic scale or to local defects in the graphene sheet, which act as nucleation sites for etching and can induce a slight shift of the channel. Etching through whole sheets and thereby forming tunnels in a direction transverse to the layers obviously is highly unfavorable.

In conclusion we showed that etching HOPG by Ni nanoparticles is not restricted to the surface, resulting in the formation of open trenches. Here we prove beyond doubt by high resolution STM and SEM imaging and triangulation experiments that tunnels are etched underneath the surface. These tunnels are frequently found and HOPG is perforated by them throughout the probing depth. Quantum chemical models confirmed that the graphene sheets forming the ceiling of the tunnels bend downward due to an attractive interaction with the tunnel walls. No rearrangement of the local electronic structure appears to be associated with this topographic change. The bending increases with increasing width of the tunnel and decreases when the number of covering graphene layers is increased.

Careful inspection of images in previous publications indicates that tunnel formation may occur also in other systems etched by Ni, Ag and Co[4-6,8,10], although it has never been explicitly reported before. This sub-surface etching-mechanism offers a novel method for fabrication of nanoporous graphite with well-defined geometric shape, which might be found useful for future applications.

## Methods

**Sample preparation.** Channels were etched by the two-step preparation method described previously[8]. Freshly peeled HOPG was exposed to air at 650 °C for 1-5 minutes. Pits with diameter up to several hundred nanometers can be created by this simple oxidation-gasification process. After oxidative etching, we deposited the catalyst by a dip-drawing process: The HOPG piece was immersed into a 0.2 mmol/L $NiCl_2$/ethanol solution, and then drawn out slowly. After drying in air, the sample was annealed at 500 °C in $Ar/H_2$ flow (1300 sccm, 15vol% hydrogen) for 1 hour. Then the temperature was immediately increase up to the cutting set-point (900 °C) and held for 1 hour.

**SEM imaging.** SEM images have been recorded with a Zeiss Ultra Plus scanning electron microscope using 10keV primary electrons and the in-lens detector that is expected to be sensitive to SE1 and SE2 secondary electrons [31]. For triangulation images have been recorded with 30keV electrons in the high-current mode using the SE2 detector. The sample was tilted relative to the electron beam around an axis which is parallel to the image´s horizontal axis. (Details in SI).

**STM investigations.** After transfer to the vacuum chamber the sample was annealed up to 150°C for 1 hour to remove water and contaminants from air. After cooling to room temperature it was transferred to our modified Omicron UHV VT STM. STM tips were electrochemically etched polycrystalline tungsten that was treated by Ar sputtering and annealing in UHV. Imaging was performed in constant current mode at RT with the voltage applied to the sample.

Linear or polynomial background was substracted from the images, Δz jumps due to tip reconfigurations have been leveled. No numerical filters have been applied to the images.

**Theoretical Modelling.** For modelling of the subsurface tunnel a slab, periodic in the zigzag <11-20> and armchair <10-10> directions was modelled. It consists of five graphene sheets of 13 periodic units $û_{<11-20>}$ along <11-20> and 2 periodic units $û_{<10-10>}$ along <10-10> direction in AB stacking order (for a detailed view see SI). Then a tunnel running along <10-10> was created by cutting out the 7.5 central units over the full width for the three inner sheets. The edge of the cut layers runs along <10-10> and corresponds to a straight armchair edge of graphene and was saturated with hydrogen atoms. This results in a tunnel with one whole bottom and top sheet. By adding one or two whole sheets on top, tunnels with two or three topping layers were created. Finally, a wider tunnel was constructed by cutting out the 10.5 central units of the inner three layers of a 5 layer stack of width 16 and depth 2 units. Technical details of the calculations are given in the supplementary information.

**Acknowledgements**

LS and PMA acknowledge the support from the office of Naval Research through the MURI program on graphene and the Exotic Nanocarbons, Japan Regional Innovation Strategy Program by Excellence, JST. PMA acknowledges the AvH foundation for the senior Humboldt-Helmoltz award to perform research in Germany. VM acknowledges the MMM@HPC EC project for financial support. ML, AV and RK thank Tobias Wassmann for fruitful discussion.

**Author Contributions**

ML, AV, PMA and RK conceived the work. LS and PMA prepared the samples. AV and RK performed the SEM, ML the STM measurements. VM performed the theoretical calculations. ML, VM, WW, RK analysed the data and prepared the manuscript. All authors discussed the data and the manuscript.

**Competing interests**

The authors have no competing interests that might be perceived to influence the results and/or discussion reported in this article.

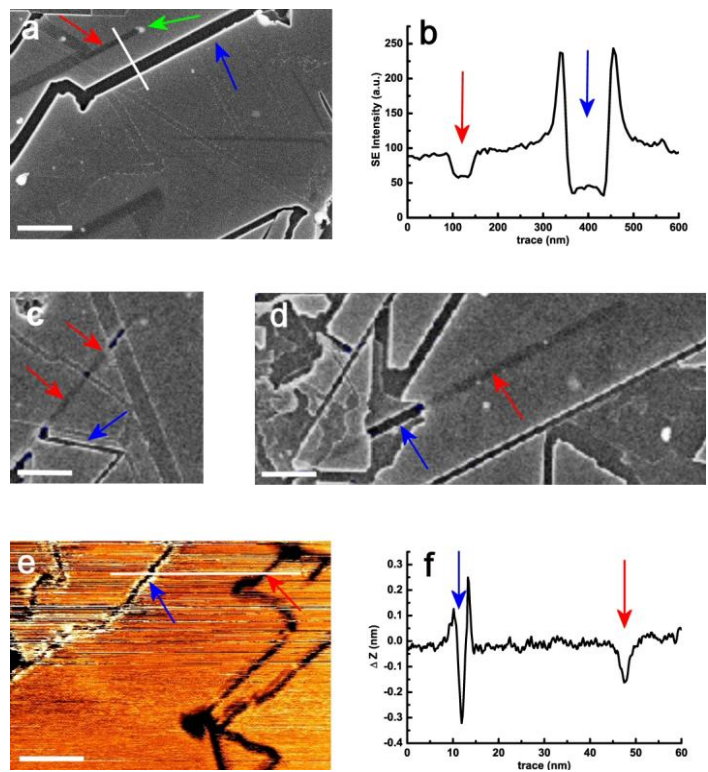

**Figure 1: Imaging of etch-channels on HOPG.** Scanning electron micrographs (a,c,d) and scanning tunneling microscopy image (e) show open channels (blue arrows), closed channels (red arrow) and sub-surface catalyst particles (green arrow). An averaged secondary electron (SE) intensity profile (b) and an apparent height profile (f) along the white lines in (a) and (e) highlight the different contrast of trenches and tunnels. Switching between trench and tunnel contrast along channels is found, where covering layers are partially removed (c,d). STM image: U=-1 V I=0.28 nA. Scalebars: 500nm (a,c,d) , 20nm (e).

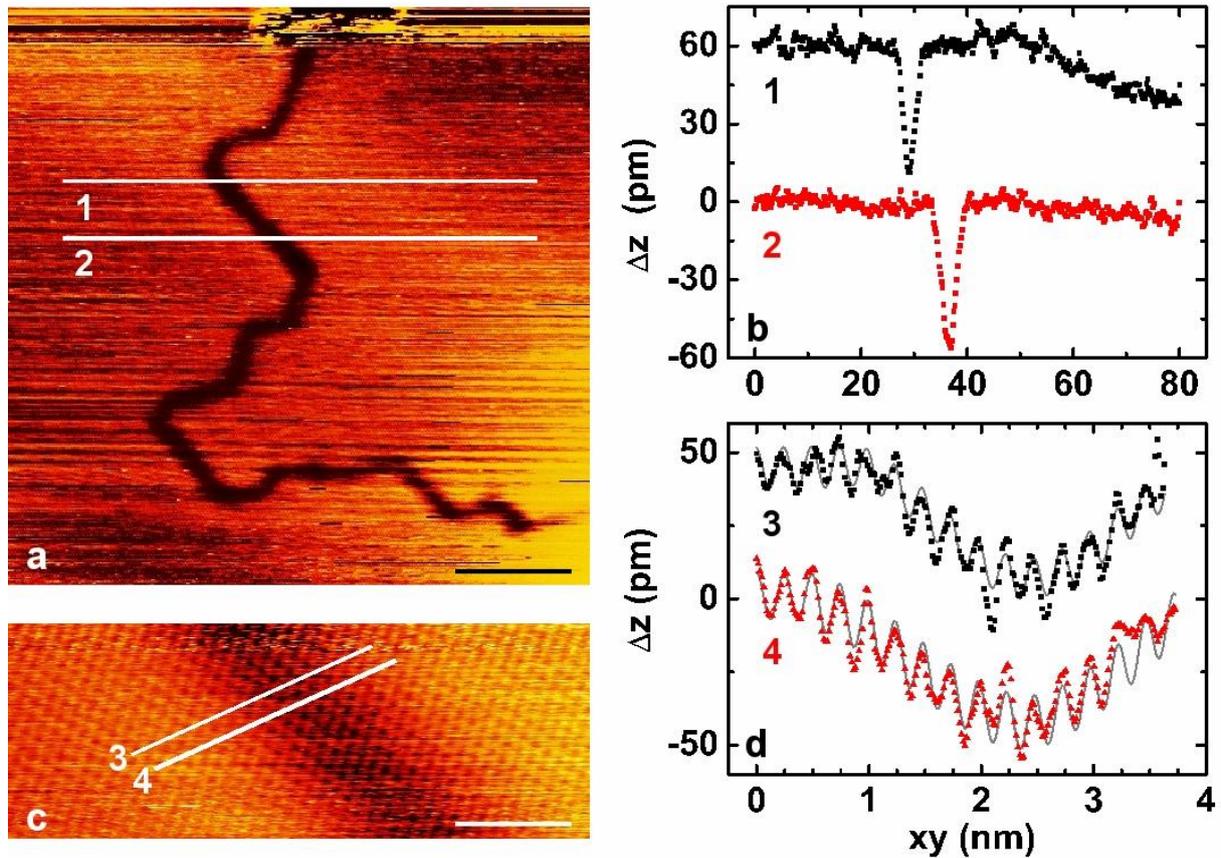

**Figure 2: High-resolution STM image (a,c) and height profiles (b,d) of a channel without edge enhancement**. The height profiles (b) of the traces marked 1,2 in (a) show a smooth sub-monolayer depression without edge enhancement. The atomically resolved STM image of the same area (c) shows an intact top graphene sheet. The height profiles (d) along the traces 3,4 in (c) show unperturbed periodicity when fitted with a sine function: Black and red symbols: measured height from STM, grey line: sine function superimposed onto third order polynomial (flexed beam model).
Scale bars: (a) 20nm, (b) 2nm, all images constant current mode, U=0,7 V, I=0.98 nA (a), I=0.74 nA (b).

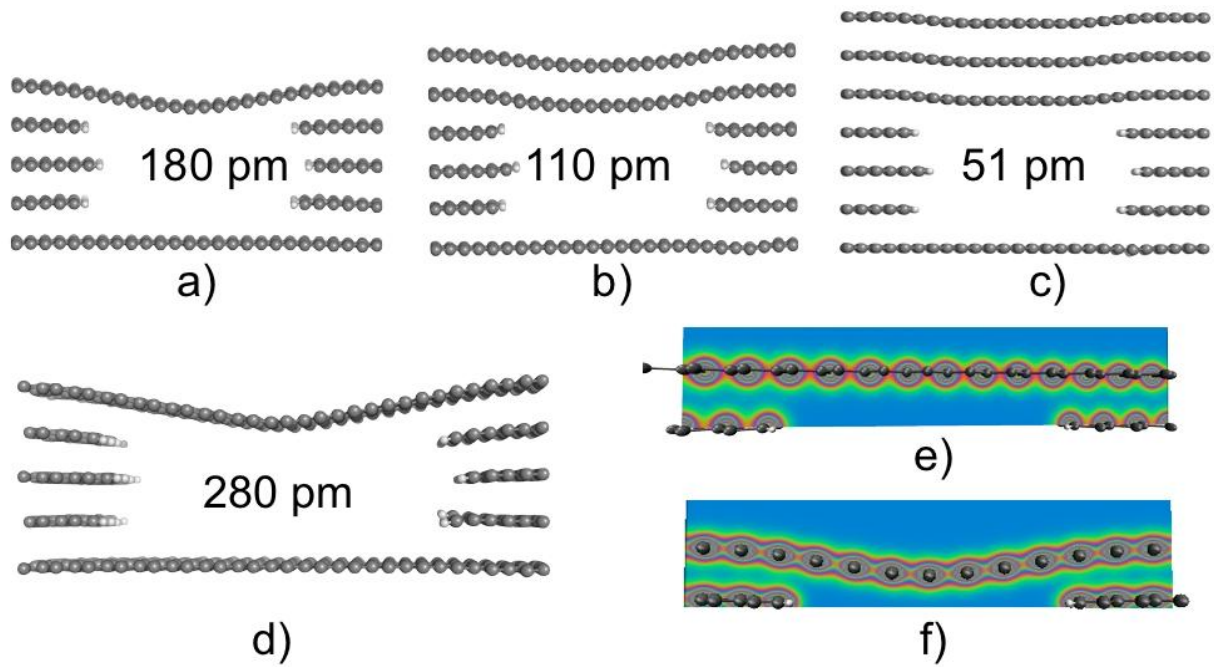

**Figure 3: Relaxed lowest energy structures (a-d) and electron density (e,f) of tunnels.** Tunnels formed by cutting 7.5 (a-c) and 10.5 (d) units $\hat{u}_{<11\text{-}20>}$ out of a stack of graphene sheets. The tunnel axis along <10-10> is running perpendicular to the paper plane. The front edge is an arm-chair edge (<11-20> running horizontally. For a top view see supplementary information). Shown structures (a-d) are the lowest energy structures after relaxation. Numbers give the maximum deviation Δz from the straight configuration. The bending of surface sheets decreases with increasing number of covering layers (a-c) and increases with tunnel width (a,d). In DFT calculations no reconfiguration of electron density is found due to the missing of layers underneath the top sheet (e) nor due to bending (f).

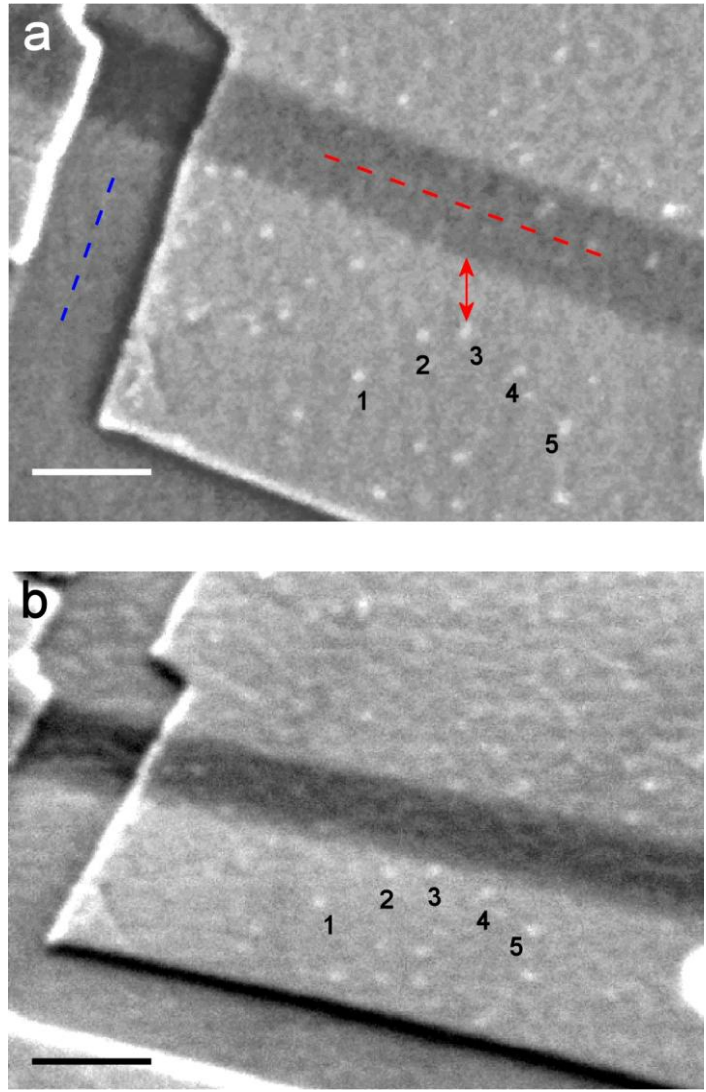

**Figure 4: Triangulation of tunnel level.** SEM micrographs of a tunnel (red dashes) and a trench (blue dashes), recorded under normal incidence of the electron beam (a) and under an angle of 45° (b) (tilt axis horizontal). The shift in the relative position between the tunnel and the surface particles (1-5) allows triangulating the vertical position of the tunnel with respect to the sample surface. The length of the red arrow translates into a depth of 75 nm. The lateral tunnel cross-section is 80 nm which corresponds to 35 nm of covering layers assuming a circular shape of the tunnel. Scalebars: 100nm.

## Supplementary Information

### Imaging of graphite and graphene in STM

It is well known that in STM-images of graphite or multilayer graphene every other atom is imaged due to the different interaction of A and B sites with the underlying graphene sheet. This results in a triangular lattice with lattice constant 0.246 nm while the nearest neighbor distance is 0.142nm[S1-S2]. Graphene on the other hand is usually imaged as a honeycomb network[S3]. Imaging of a single graphene layer by a triangular lattice may happen due to relaxations of every other atom due to anisotropic stress[S4]. At our feedback parameters we do not observe indications for a manipulation of the surface sheet by the tip resulting in such stress. We also do not find an anisotropic buckling within the structure in our calculations of relaxed tunnel cover-layers. Since we never observe an unambigous honeycomb lattice we therefore assume that the tunnels are covered by two or more layers of graphene.

### Theoretical modeling

Due to the size of the samples in question the structural optimization was performed with modified neglect of diatomic differential overlap (MNDO) a semi–empirical quantum chemistry method with PM6-DH2[S5] parametrization as implemented in MOPAC[S6]. The method was tested for graphite bulk resulting in an in-plane carbon-carbon distance of 1.42 Å and 3.34 Å inter planar distance, in excellent agreement with experiment.

The tunnel is modeled by large supercells (see fig. S1) with lattice vectors along zigzag <11-20> and armchair <10-10> directions. A unit cell along these directions is marked in green in fig S1. The super cell for Fig. 3 (a-c) is constructed by an AB stack of sheets of 13 unit cells along $\hat{u}_{<11-20>}$ and 2 unit cells along $\hat{u}_{<10-10>}$. The tunnel is constructed by cutting out 7.5 unit cells, this means 15 rows of atoms – oriented along $\hat{u}_{<1o-10>}$- are missing. Fig. S1 shows the three lowest layers of the super cell looking down the <0001> direction. Hydrogen atoms are not shown. The lowest layer (black) corresponds to an intact graphene layer. The next two layers as well as the forth layer (not shown, structure as 2$^{nd}$ layer) form the side walls of the tunnel. The armchair edge is indicated as grey line for the 3$^{rd}$ layer. The cover sheet (5$^{th}$ layer, not shown) is equivalent to the bottom sheet. For Fig. 3 (b) and (c) one or two additional sheets have been added on top, respectively. The wider tunnel (Fig. 3d) is constructed accordingly with 10.5 unit cells (21 rows) cut out of a stack of 16 $\hat{u}_{<11-20>}$ width. Three layers form the tunnel´s side wall; one sheet is covering the tunnel. The geometry optimization was performed on all atoms. For the top layers the straight, symmetric configuration was broken by hand. The bottom is kept symmetric reflecting the bulk graphite geometry assumed to be below.

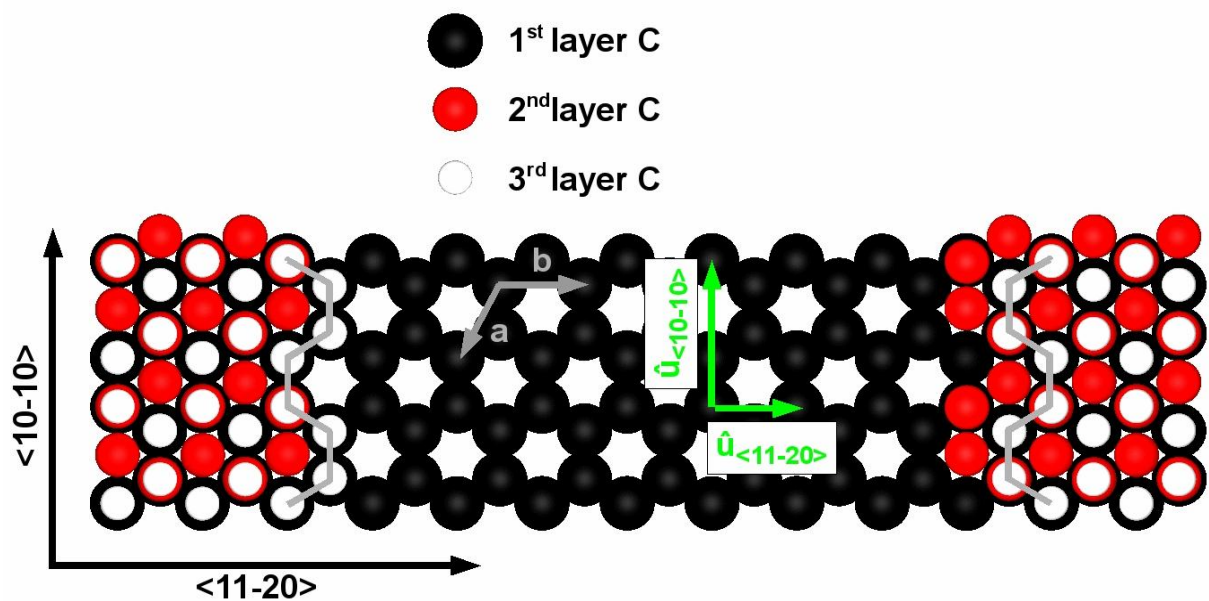

**Figure S1: Construction of the narrow tunnel.** Carbon atoms of the bottom (black), second (red) and third (white) layer are shown in different sizes, hydrogen atoms are not shown for visibility. The unit cell for the miller indices [a,b,-(a+b),z] is given in grey. The unit vectors û as referred to in the text are indicated green. The armchair edge of the side wall of the tunnel is marked by a grey line for the 3$^{rd}$ layer.

The single point DFT calculations were performed on structures consisting of one infinite graphene sheet on top and a side-wall sheet below. Two structures were considered one straight and one bent as the MOPAC optimized structure with the top being formed by a single graphene layer. For this purpose we used the Vienna *ab-initio* simulation package (VASP)[S7]. The package utilizes a plane-wave basis[S8] and projector augmented wave (PAW) pseudo-potentials[S9] within the generalized gradient approximation (GGA)[S10] (comparison with local density approximation – LDA gave no significant difference). We used a Monk-horst 5x9x1 k point grid[S11]. In the z direction we added 10 Å of vacuum in order to prevent any cross-talk between the slabs.

**Triangulation of the depth-level of the tunnel**

When the sample is tilted, the distance between points on the same height level decreases due to projection. The distance between objects on different height level may as well change their relative position.

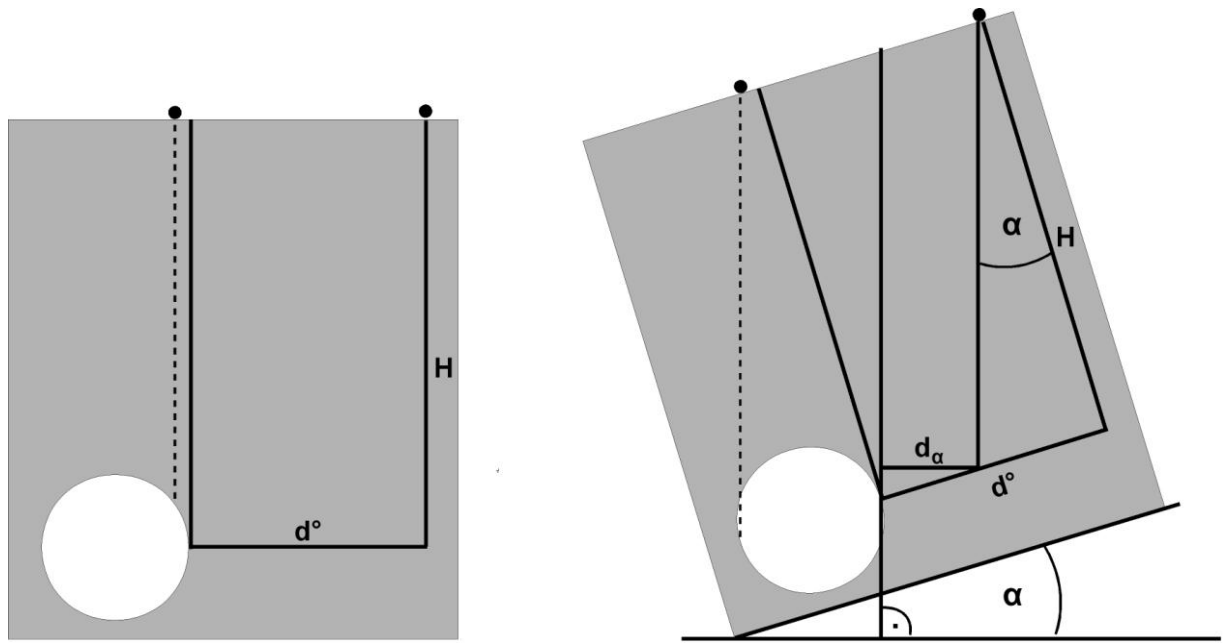

**Figure S2: Triangulation of depth-level H.** Apparent distance between particle at surface and side wall of the tunnel changes from $d°$ to $d_\alpha$ with tilt angle $\alpha$. For objects on the same level (e.g. two particles on top) the distance becomes smaller while the relative position stays the same at all angles. Objects on different height levels may change their relative position with angle: the left particle on the surface appears to change from "the right side" of the tunnel to the "left".

The surface particle on the left in Fig. S2 shifts from "at the right" to "at the left" of the tunnel. This is observed in Fig. 4 (main text) as some of the small particles shift from "on top" of the tunnel to at the side ("above" in image b) of the tunnel, others from at the side ("below" in image a) to "on top". The tunnel also slips past the kink of the trench edge in the upper left corner of Fig. 4. From the shift of a surface object relative to the tunnel edge we can estimate the depth-level of the tunnel below the surface to be $H = (d° - d_\alpha \cos \alpha) / \tan \alpha$.